\documentclass[twocolumn,english,prl,showpacs,superscriptaddress,aps]{revtex4-1}
\usepackage[T1]{fontenc}
\usepackage[latin9]{inputenc}
\usepackage[nointlimits]{amsmath}
\usepackage{amssymb}
\usepackage{graphicx}
\usepackage{esint}
\usepackage{times}

\newcommand{\LGPE}{\mathcal{L}}

\makeatletter
\@ifundefined{textcolor}{}
{%
 \definecolor{BLACK}{gray}{0}
 \definecolor{WHITE}{gray}{1}
 \definecolor{RED}{rgb}{1,0,0}
 \definecolor{GREEN}{rgb}{0,1,0}
 \definecolor{BLUE}{rgb}{0,0,1}
 \definecolor{CYAN}{cmyk}{1,0,0,0}
 \definecolor{MAGENTA}{cmyk}{0,1,0,0}
 \definecolor{YELLOW}{cmyk}{0,0,1,0}
 }

\@ifundefined{definecolor}{\@ifundefined{definecolor}
 {\@ifundefined{definecolor}
 {\@ifundefined{definecolor}
 {\@ifundefined{definecolor}
 {\@ifundefined{definecolor}
 {\usepackage{color}}{}
}{}
}{}
}{}
}{}
}{}
\makeatother

\usepackage{babel}
\begin{document}
\title{Yang-Yang thermometry and momentum distribution of a  trapped one-dimensional Bose gas}

\author{M.~J.~Davis}
\altaffiliation[]{These authors contributed equally to this work.}
\affiliation{The University of Queensland, School of Mathematics and Physics, Qld 4072, Australia}
\author{P.~B.~Blakie}
\altaffiliation[]{These authors contributed equally to this work.}
\affiliation{Jack Dodd Centre for Quantum Technology, Department of Physics,  University of Otago,  Dunedin, New Zealand}
\author{A.~H.~van~Amerongen}
\altaffiliation[Current address: ]{SRON, Netherlands Institute for Space Research,  Utrecht, The Netherlands.}
\affiliation{Van der Waals-Zeeman Institute, University of Amsterdam, Science Park 904, 1098 XH Amsterdam, The Netherlands}
\author{N.~J.~van~Druten}
\affiliation{Van der Waals-Zeeman Institute, University of Amsterdam, Science Park 904, 1098 XH Amsterdam, The Netherlands}
\author{K.~V.~Kheruntsyan}
\affiliation{The University of Queensland, School of Mathematics and Physics, Qld 4072, Australia}

\date{\today}
\begin{abstract}
We describe the use of the exact Yang-Yang solutions for the one-dimensional
Bose gas to enable accurate kinetic-energy thermometry based on
the root-mean-square width of an experimentally measured momentum distribution. Furthermore, we use the
stochastic projected Gross-Pitaevskii theory to provide the first quantitative description of the full momentum distribution
measurements of Van Amerongen et al., Phys. Rev. Lett. \textbf{100}, 090402 (2008). We find the fitted
temperatures from the stochastic projected Gross-Pitaevskii approach are
in excellent agreement with those determined by Yang-Yang kinetic-energy thermometry.
\end{abstract}


\maketitle
Ultracold gases offer a unique opportunity to study
fundamental problems in quantum many-body physics, allowing
 experimental observations to be compared directly with microscopic theories.
  An area of significant recent interest has been the measurement of thermodynamic  relations  
 \cite{,Tan-a,*Braaten08,*Leggett09,*Ho-2010,NAS10,*Vale10,*Tung2010a,*Rath2010a,*Hung2011a,ARM10b}.
The one-dimensional (1D) Bose gas with repulsive interactions 
has  emerged as a paradigm system because exact solutions are
available for both eigenstates  \cite{LL} and
thermodynamic quantities \cite{YY} (see Ref.~\cite{BouchouleChipBook}
 and references therein).
Furthermore, this system exhibits a surprisingly rich variety of regimes \cite{KK03,KK05} connected by broad crossovers. 
Most studies of the 1D Bose gas have focused on the position-space distributions \cite{KK05,Bouchoule,Amerongen,ARM10b} 
and  local correlations 
\cite{Gangardt-2003,KK03,KK05,Cazalilla-2004,Mora-Castin,Zvonarev-2006}, which can be directly obtained from the exact theories.

A recent experiment by Van Amerongen \textit{et al.}~\cite{Amerongen} measured the position and momentum distributions of a trapped 
1D Bose gas   throughout the crossover from an ideal gas to the quasicondensate regime.   The  position-space measurements were compared with the 
Yang-Yang (YY) thermodynamic solutions~\cite{YY}
within the local density approximation (LDA), and showed  smooth behavior throughout the crossover. In contrast, the momentum distributions showed a  pronounced temperature dependence,  and have been
unexplained by theory
to date.  Previous work on the momentum properties of the 1D Bose gas has focused on limiting cases
\cite{Mora-Castin,Cazalilla-2004,Minguzzi2002a,*Lapeyre2002a,*Olshanii-2003,*Gerbier-2003,*Deuar-2004}.

Here we investigate the momentum properties of the 1D Bose gas and their application to thermometry through measurements of the system kinetic energy.  
Our methods provide a reliable foundation for accurate thermometry in
all regimes of a 1D Bose gas with repulsive interactions, including  the strongly-correlated
regime. This approach is  reminiscent of molecular dynamics calculations, 
where the average kinetic energy per particle  is a direct measure of the temperature  \cite{Haile}.

First, we use the exact YY thermodynamic formalism \cite{YY} to calculate the root-mean-square (rms) width of the momentum distribution, 
which is equivalent to determining the average kinetic energy per particle. In combination with the LDA for  trapped 
(nonuniform) quasi-1D systems, we show how the  YY kinetic energy results can be applied to accurate thermometry for
a broad range of conditions that are relevant to ongoing experimental and theoretical efforts. We refer to this approach as YY thermometry.
Second, we present the first  quantitative calculation of the full momentum distribution  for a trapped quasi-1D Bose gas using the stochastic projected Gross-Pitaevskii equation (SPGPE) technique \cite{Gardiner2003a,*cfieldRev2008}. (Calculations of the position-space distribution based on a related formalism \cite{Cockburn2011a} were recently described in \cite{Cockburn2011b}.)
We find excellent agreement between the SPGPE results and the momentum-space measurements reported in Ref.~\cite{Amerongen}. Finally, we compare the kinetic energy and temperature predictions of the YY thermometry with those of the SPGPE and also find excellent agreement between the two approaches.
While the SPGPE technique is limited to the degenerate yet high-temperature weakly interacting regime, 
the YY thermodynamic formalism applies to all repulsive interaction strengths and temperatures and therefore its usefulness extends beyond the regimes studied 
here.

A uniform 1D Bose gas in the thermodynamic limit is completely characterized
by two parameters \cite{YY,LL,KK03,KK05,Bouchoule,BouchouleChipBook}: the dimensionless interaction strength $\gamma=mg/\hbar^{2}\rho$
and the reduced temperature $t\equiv2k_{B}T\hbar^{2}/mg^{2}$,
where  $\rho$ is the linear (1D)
density,  $g\simeq2\hbar\omega_{\perp}a$ is the effective 1D coupling strength \cite{olshanii-1d-scattering},
 $a$  is the 3D $s$-wave scattering length, and $\omega_{\perp}$
the transverse radial harmonic trapping frequency
\footnote{This 
expression for $g$ assumes that $a\ll l_{\perp}\!=\!\sqrt{\hbar/m\omega_{\perp}}$ \cite{olshanii-1d-scattering}}.
The 1D regime is realized when the transverse excitation energy $\hbar\omega_{\perp}$
is much larger than the thermal energy $k_{B}T$ and  chemical
potential $\mu$. While the YY thermodynamic equations do not directly yield 
the 1D momentum distribution, $n(k_z)$, here we show  how they can be used to obtain the average kinetic energy per particle
\begin{equation}
E_{\mathrm{kin}}/N = \hbar^{2}\langle k_z^2\rangle/2m,
\label{E-nk}
\end{equation}
where 
$\langle k_z^2\rangle^{1/2}\!=\![\int \!dk_z\,k_z^2\,n(k_z)/N]^{1/2}$ is the  rms width of $n(k_z)$, and $N\!=\!\int \!dk_z n(k_z)$ is the total atom number.

Solutions to the YY thermodynamic equations yield a unique value for the total energy per particle, $E/N$ \cite{YY}, for a given temperature and interaction strength. Using the Helmann-Feynman theorem, the YY solutions can also be used to determine the local pair correlation function $g^{(2)}(0)$ \cite{KK03,KK05}.  This gives the interaction energy per particle, $E_{\mathrm{int}}/N=\frac{1}{2}g\rho g^{(2)}(0)$, and thus the kinetic energy per particle 
is  found from $E_{\mathrm{kin}}/N=E/N-E_{\mathrm{int}}/N$.

\begin{figure}
\includegraphics[width=6.9cm]{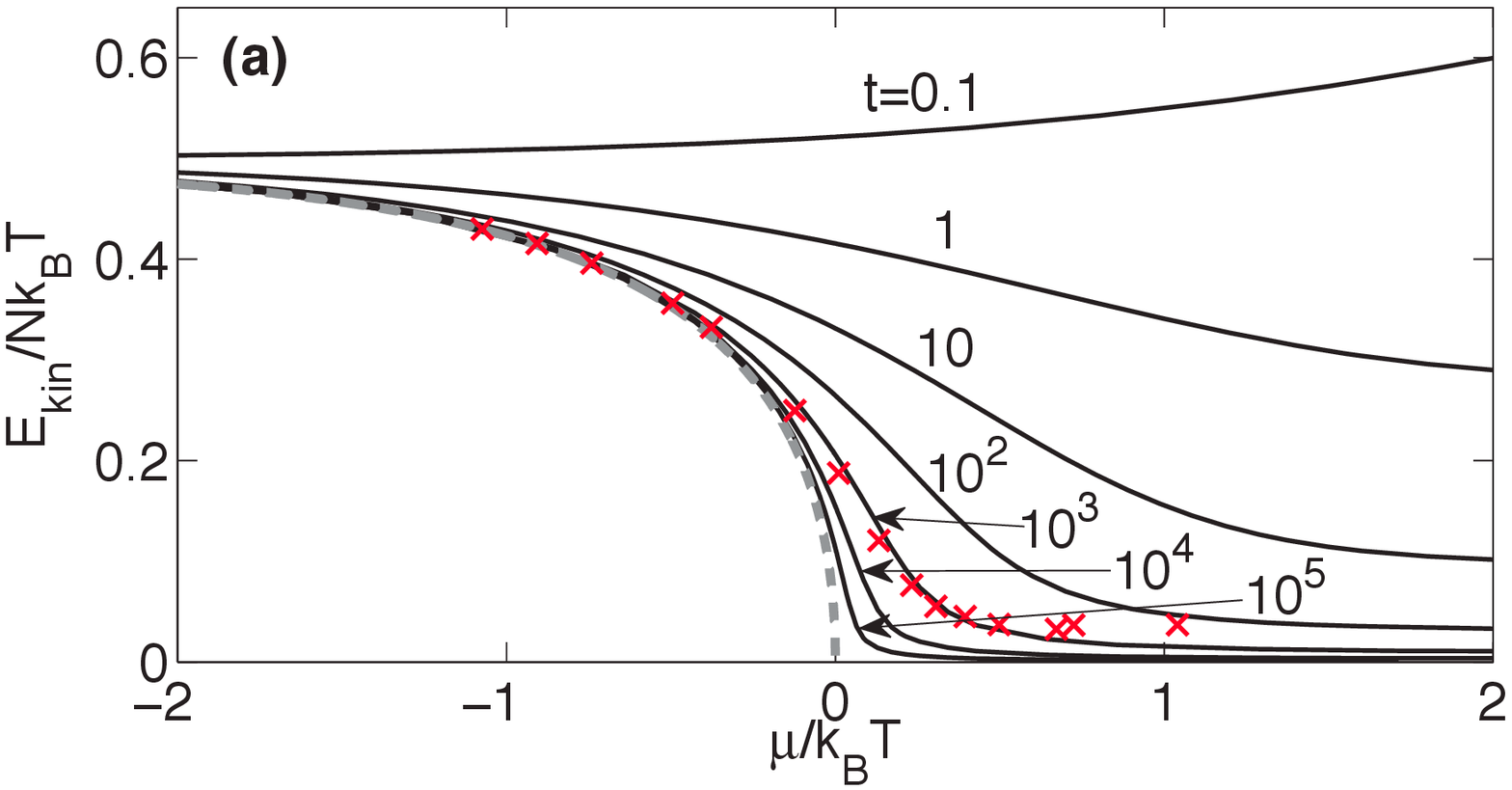}
\includegraphics[width=6.9cm]{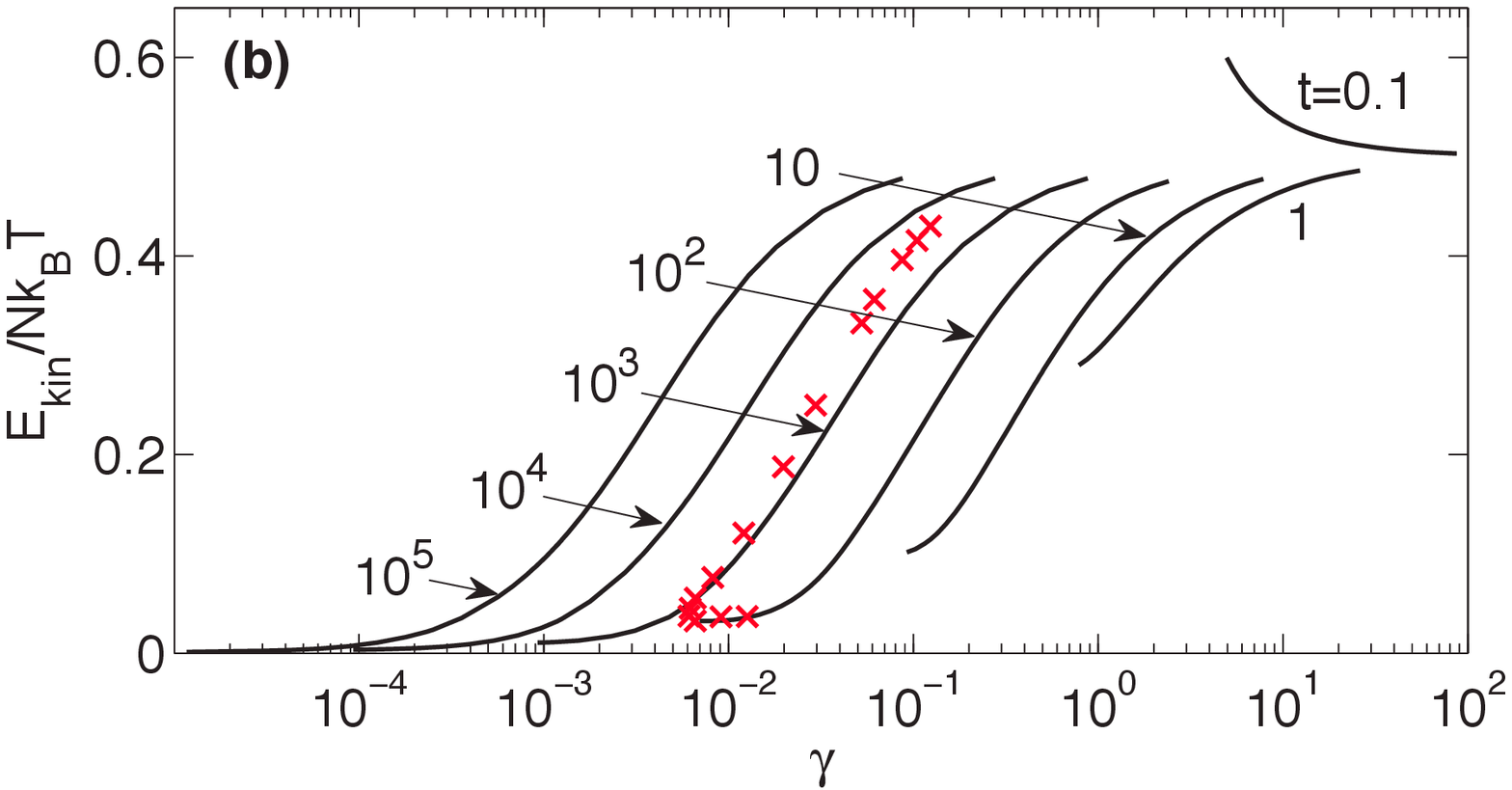}
\caption{(Color online) (a) Solid lines show the kinetic energy per particle, $E_{\mathrm{kin}}/N$ (in units of $k_BT$), from the exact YY solutions for a
uniform 1D Bose gas as a function of the chemical potential $\mu$ for different values of the dimensionless 
temperature parameter $t$. The gray dashed line is the ideal-Bose-gas result,
corresponding to $t\!\rightarrow\!\infty$ ($\gamma\!\rightarrow \!0$). 
The crosses show the local kinetic energy per particle  (without the contribution of the excited transverse modes)  in the center
of the trap 
for the experimental data presented in 
Fig.~\protect\ref{fig:Ekin}. (b) Same data
as in (a), i.e., within the same window of values of $\mu/k_{B}T$,
but presented as a function of the dimensionless 
interaction parameter
$\gamma$ using the exact equation of state $\rho\!=\!\rho(\mu,T)$ \cite{Bouchoule}. 
}
\label{fig:Ekin_from_YY} 
\end{figure}

The exact kinetic energy per particle for the uniform system, calculated for a wide range of interaction strengths and  temperatures, is
shown in Fig.~\ref{fig:Ekin_from_YY}.
These results reveal that $E_{\mathrm{kin}}/N$
varies considerably as the degeneracy and  interaction strength are changed; thus  it is a useful observable for discriminating between different  regimes of the 1D Bose gas \cite{KK03,KK05,BouchouleChipBook}. We identify three broad regimes of behavior. 
(\textit{i}) 
For $\mu<0$ and $|\mu|/k_{B}T\gg1$, the system behaves as a \textit{classical} (nondegenerate) ideal gas with $E_{\mathrm{kin}}/Nk_{B}T\!\to\!1/2$, as expected from the classical equipartition theorem. As 
$\mu\! \rightarrow \! 0^{-}$, the kinetic energy per particle, for sufficiently large values of $t$ (negligible interactions), decreases as expected for a \emph{degenerate} ideal 1D Bose gas [gray dashed line in Fig.~\ref{fig:Ekin_from_YY}(a)],  
${E_{\mathrm{kin}}}/{Nk_{B}T} \! = \! \frac{1}{2} {g_{3/2}(\lambda)}/{g_{1/2}(\lambda)}$,
where $g_{s}(\lambda)\!=\!\sum_{l=1}^{\infty}\lambda^{l}/l^{s}$ is a Bose function and $\lambda\!=\!e^{\mu/k_{B}T}$.
(\textit{ii}) 
Once $\mu$ becomes positive, the degenerate behavior is strongly affected by interactions. 
For $t\!\gg\!1$ (weak interactions, $\gamma\!\ll\!1$) the system is a quasicondensate 
and there is  a significant reduction of $E_{\mathrm{kin}}/N$ from the
classical equipartition value.
 (\textit{iii}) 
 For $t<1$ (strong interactions, $\gamma>1$) the system becomes fermionized, leading to an increase in
$E_{\mathrm{kin}}/N$ for both negative and positive $\mu$. For $t\ll1$, and $1\ll\gamma\ll t^{-1/2}$~\cite{KK05} in the Tonks-Girardeau regime, $E_{\mathrm{kin}}/Nk_BT$ becomes  larger than the equipartition value of $1/2$ \footnote{In this regime $E_{\rm{kin}}/Nk_BT = \mu/3k_BT=\pi^2/6 t \gamma^2$, with $\mu=\pi^2 \hbar^2 \rho^2 /2m$.}.

In order to apply the YY results 
to trapped (nonuniform) quasi-1D systems as realized in experiment~\cite{Amerongen}, 
we utilise the LDA.
The trapping potential is  $U(\mathbf{r})\!=\!V(z)\!+\!\frac{1}{2}m\omega_{\perp}^{2}(x^{2}+y^{2})$,
where $V(z)\!\simeq\! \frac{1}{2}m\omega_{z}^{2}z^{2}$ is the potential 
in the weakly confined 
longitudinal direction with $\omega_{z}\ll\omega_{\perp}$ \cite{anharmonicity}.
We calculate the particle
number density $\rho(z)$ by treating the trapped system as a collection
of sufficiently small uniform systems of length $\Delta z$,
with the local chemical potential  {$\mu(z)=\mu-V(z)$} \cite{KK05,Bouchoule}
where $\mu$ is the global chemical  potential.
The density of
the trapped system in the ground  transverse mode, $\rho_{0}(z)$,
is evaluated as $\rho_{0}(z)=\rho_{\mathrm{YY}}[\mu(z),T]$, where $\rho_{\mathrm{YY}}[\mu,T]$
is the YY density for a uniform  system.  Similarly, the kinetic energy of the ground  transverse
mode is 
\begin{equation}
E_{\mathrm{{kin},0}}=\int dz\,\mathcal{E}_{\mathrm{kin}}[\mu(z),T],
\end{equation}
where $\mathcal{E}_{\mathrm{kin}}=E_{\mathrm{kin}}/\Delta z$ is the kinetic energy
density of a uniform system of length $\Delta z$.

In the experiment \cite{Amerongen} the system temperature was not sufficiently low that all transverse excitations were frozen out.  
We can account for this by observing that  $g\rho_0\ll\hbar\omega_{\perp}$ so that the transverse
excitations 
are well approximated as harmonic oscillator
states with energies $j\hbar\omega_{\perp}$ ($j=0,1,2,...$), 
where we have removed the zero-point energy $\hbar\omega_{\perp}$.
The transversely excited states can then be accurately described as independent ideal 1D Bose gases
 with chemical potentials $\mu_j(z)=\mu(z)-j\hbar\omega_{\perp}$ \cite{Amerongen}. 
Accounting for the degeneracy factor $j+1$,
the  1D position- and momentum-space densities for the atoms in transversely excited states are,  respectively,
 \begin{eqnarray}
\rho_{{\rm {e}}}(z) &=&\sum_{j=1}^{\infty}\frac{j+1}{\Lambda_{T}}g_{1/2}\left[e^{\beta      \mu_j(z)} \right]\!,\label{rho-e-x-1}\\
n_{\rm{e}}(k_z) &=&\sum_{j=1}^{\infty}(j+1)\!\int \frac{dz }{2\pi}\left\{e^{\beta\left[\frac{\hbar^2k_z^2}{2m}-\mu_j(z)\right]}-1\right\}^{-1}\!,\label{rho-e-k-1}
\end{eqnarray}
where $\beta=1/k_BT$, 
and $\Lambda_{T}=(2\pi\hbar^2/mk_BT)^{1/2}$ is the thermal de Broglie wavelength. 
Using Eq. (\ref{rho-e-k-1}) we  obtain  the kinetic energy of the excited modes, $E_{\mathrm{kin,e}}$ [cf.~Eq.~(\ref{E-nk})].
Combining these results, the full YY description of the trapped quasi-1D
system is given by the total density, $\rho(z)=\rho_{0}(z)+\rho_{\rm e}(z)$,
total atom number $N\!=\!\int\! \rho(z) dz$, and the total average kinetic energy  per particle
 ${E_{\mathrm{kin,t}}}/N= (E_{\mathrm{kin,0}}+E_{\mathrm{kin,e}} )/N$.

\begin{figure}[t]
\includegraphics[width=8.6cm]{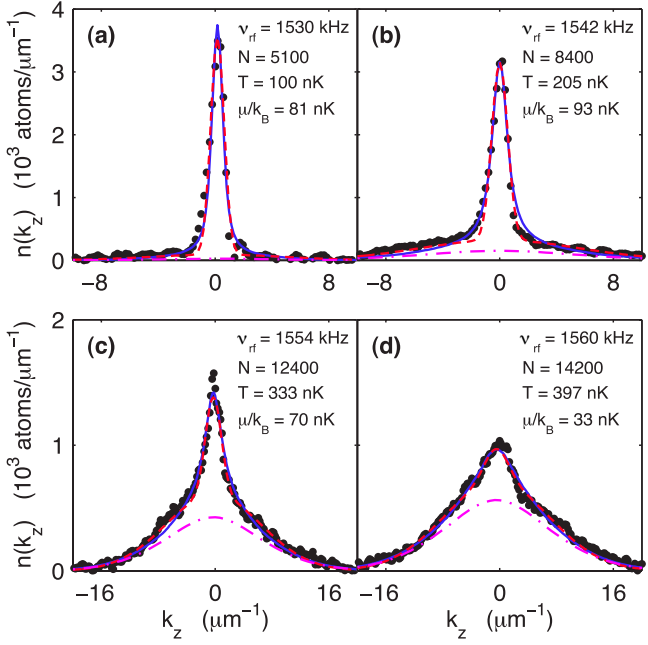}
\caption{(Color online) Examples of the experimental momentum distributions (black dots), bimodal Gaussian fits (red dashed lines)  and theoretical SPGPE best-fits (blue solid lines) yielding the temperature $T$ { and kinetic energy $E_{\mathrm{kin,t}}$.} The bimodal Gaussian fits are difficult to distinguish from the SPGPE fits, but give slightly different values for the total kinetic energy (see inset of Fig.~\ref{fig:Ekin}). The magenta dot-dashed lines indicate the density of atoms in transversely excited modes $n_{\mathrm{e}}(k_z)$.
The finite experimental imaging resolution was modeled as a Gaussian with an rms width of $2.83$~$\mu$m~\cite{Amerongen,Amerongen-thesis}.  The uncertainty in the stated values of $T$ and $\mu/k_B$ is typically $\pm 5$ nK (95\% confidence interval). The values of $\nu_{\mathrm{rf}}$ correspnd to final rf frequency for evaporative cooling.}
\label{fig:nk-fits} 
\end{figure}

 Before comparing the YY thermodynamic predictions for $E_{\mathrm{kin,t}}/N$ with the 
 experimental measurements of Ref.~\cite{Amerongen}, 
 we outline the second theoretical approach used here---the SPGPE---which allows for the determination of both global 
 thermodynamic quantities and the full momentum distribution in the weakly interacting regime.
In the SPGPE approach, the system
field operator for the lowest transverse mode is split into two parts, $\hat{\Psi}_0(z)=\hat{\psi}_{\mathrm{C}}(z)+\hat{\psi}_{\mathrm{I}}(z)$,
representing coherent 
($\hat{\psi}_{\mathrm{C}}$) and 
incoherent ($\hat{\psi}_{\mathrm{I}}$) regions \cite{Gardiner2003a,*cfieldRev2008}. The coherent region is defined by an energy  cutoff $\epsilon_{\mathrm{cut}}$ such that it contains all highly occupied  modes.  It can then be described as a classical field  (i.e.~$\hat{\psi}_{\mathrm{C}}\to{\psi}_{\mathrm{C}}$) evolving according to the simple growth SPGPE  \cite{Gardiner2003a,*cfieldRev2008},
\begin{equation}
d\psi_{\mathrm{C}}=\mathcal{P}\!\left\{\!\left(\frac{\gamma_d}{k_BT}-\frac{i}{\hbar} \right)\!(\mu-\LGPE)\psi_{\mathrm{C}} dt+\sqrt{2\gamma_d}dW\right\}\!,\label{uSPGPE}
\end{equation}
where
$\LGPE\!=\!-\frac{\hbar^2}{2m}\partial_z^{2} \!+\!V(z)\!+\!g|\psi_{\mathrm{C}}(z)|^2$ is the Gross-Pitaevskii operator,  the parameters 
$\gamma_d$, $\mu$, and $T$ are the damping rate, chemical potential, and temperature of the reservoir, respectively, and  $dW$ is a complex Gaussian noise that is delta-correlated in time and space. This equation explicitly includes a  projection ($\mathcal{P}$) onto the coherent-region modes \cite{Blakie2005a},  and
can be derived  from a microscopic theory by tracing out the high energy modes that act as a thermal reservoir  \cite{Gardiner2003a,*cfieldRev2008}. In steady-state evolution the SPGPE samples $\psi_{\mathrm{C}}$  from a grand canonical density \emph{independent} of the value of $\gamma_d$. Thus the equilibrium density can be sampled in both position [$\rho_{\mathrm{C}}(z)=\overline{|\psi_{\mathrm{C}}(z)|^2}$] and momentum [$n_{\mathrm{C}}(k_z)=\overline{|\phi_{\mathrm{C}}(k_z)|^2}$] space, where the overline indicates time-averaging, and 
$\phi_{\mathrm{C}}(k_z)$ is the spatial Fourier transform of $\psi_{\mathrm{C}}(z)$.

The incoherent region (i.e., the longitudinal states of the ground transverse mode with low occupation) is well-described using the Hartree-Fock approximation  \footnote{
This is consistent with the conclusions of J.-B. Trebbia \emph{et al.}, Phys.~Rev.~Lett.~\textbf{97}, 250403 (2006), who found that the Hartree-Fock
theory failed to describe the \emph{entire} 1D Bose gas, i.e.,~the combination of both the coherent and incoherent regions.}, and has position density
\begin{equation}
\rho_{\mathrm{I}}(z)=\int_{\epsilon_{k_z}^0>\epsilon_{\rm{cut}}}{\frac{dk_z}{2\pi}}\frac{1}{e^{\beta[\epsilon_{k_z}^0+2g\rho_{\mathrm{C}}(z)-\mu]}-1  },
\end{equation}
where 
$\epsilon_{k_z}^0=\hbar^2k_{z}^{2}/2m+V(z)$ \cite{anharmonicity}.
The incoherent region 
 momentum density $n_{\mathrm{I}}(k_z)$ is obtained by a similar procedure,  and  the transverse ground mode distribution is  $n_0(k_z)=n_{\mathrm{C}}(k_z)+n_{\mathrm{I}}(k_z)$. Atoms in the 
excited transverse 
 modes are treated as for the YY formalism, thus giving the full
 momentum distribution and total kinetic energy.

Examples of best-fit momentum distributions obtained using the SPGPE approach 
{\footnote{{The best-fit SPGPE momentum distributions were determined by calculating the function $n(k_z,\mu,T)$, and performing a least-squares fit  to the experimental data constrained to the experimentally determined atom number.}}}
are compared to the experimental data 
{\footnote{The experimental momentum distributions were 
obtained by applying a focusing pulse to the trapped
atoms, followed by a rapid switch off of the trapping potential~\cite{Amerongen}. The
cloud then rapidly expands in the radial direction, effectively switching off the interactions, while it is still contracting axially. 
The axial contraction can therefore be treated as free propagation so that the measured spatial distribution
in the focus is converted to a momentum distribution via the time-of-flight transformation $\hbar k_z = m z / \tau$.
The timescale $\tau$ is calculated
using the ABCD matrix formalism for matter-wave propagation 
giving $\tau = 15$~ms with an estimated systematic uncertainty of $10$\% 
[for a detailed discussion, see Ref.~\cite{Amerongen-thesis}, in particular Section 4;  see also
J.-F. Riou \textit{et al.}, Phys. Rev. A \textbf{77}, 033630 (2008)]. A very similar
method was used to obtain the momentum distribution of a (nearly) 2D
Bose gas: S. Tung, G. Lamporesi, D. Lobser, L. Xia, E. A. Cornell, Phys.
Rev. Lett.  \textbf{105}, 230408 (2010).} 
in Fig.~\ref{fig:nk-fits}. We find quantitative agreement throughout the crossover from the nearly 
ideal Bose gas to the weakly interacting  quasicondensate regime. 
From these fits we can  determine both the kinetic energy per particle and the 
temperature for the data.

\begin{figure}[t]
\includegraphics[width=7cm]{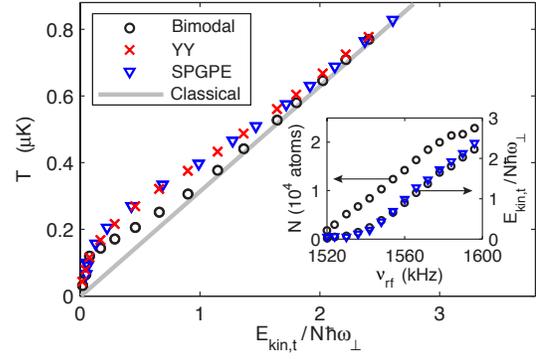}  
\caption{(Color online) Temperature of a quasi-1D Bose gas in a harmonic
trap 
(with $\omega_z/2\pi=8.5$~Hz and $\omega_{\perp}/2\pi={3280}$~Hz~\cite{anharmonicity})
as a function of the total kinetic energy $E_{\mathrm{kin,t}}$ determined by bimodal Gaussian fits (circles)~\cite{Amerongen},  SPGPE fits (triangles) \cite{HF}, YY thermometry (crosses), and the classical ideal gas model (gray line). The inset shows the atom number (open circles, left axis) and $E_{\mathrm{kin,t}}$ (right axis) as determined by bimodal Gaussian fits (open circles) \cite{Amerongen} and SPGPE fits (triangles) to the  momentum distributions as a function of the final rf evaporation frequency $\nu_{\mathrm{rf}}$.  }
\label{fig:Ekin} 
\end{figure}

In Ref.~\cite{Amerongen} the experimental momentum distributions were fitted with the sum of two Gaussians (see Fig.~\ref{fig:nk-fits}), giving the total number of atoms $N$, and  
the total kinetic energy $E_{\mathrm{kin,t}}$. The temperature $T$ was obtained from the width of the broadest Gaussian component using a classical ideal-gas model.    {We note that there is very little quantitative difference between the heuristic bimodal Gaussian momentum fits and those based on the microscopic SPGPE formalism.  However, there is a distinct difference in the temperatures extracted using the two methods (see Fig.~\ref{fig:Ekin}).

We} can also use the YY formalism for thermometry in this system, as there is a  one-to-one correspondence between the temperature $T$ and the kinetic energy per particle $E_{\mathrm{kin,t}}/N$ for a given atom number $N$.  We compare the temperature estimates of YY thermometry, the SPGPE momentum fits~\cite{HF}, and the broad Gaussian fits as a function of $E_{\mathrm{kin,t}}/N$ in { the main panel of }Fig.~\ref{fig:Ekin}.  A key result of this paper is that the YY temperatures collapse to the same curve as that obtained from the SPGPE fits to the full momentum distributions.  }

{ The inset of Fig.~\ref{fig:Ekin} shows the total number of atoms as a function of the final rf frequency for evaporative cooling, as well as a comparison of $E_{\mathrm{kin,t}}$ from the bimodal Gaussian fits and the SPGPE momentum fits. }The temperature estimates from the SPGPE and YY methods  differ for each final $\omega_{\mathrm{rf}}$ due to small differences between the SPGPE and bimodal Gaussian fits (see the comparison in Fig.~\ref{fig:nk-fits}).  However, if the YY thermometry utilised the values of $E_{\mathrm{kin,t}}$ determined from the SPGPE fits, the temperature estimates would be \emph{identical}. Thus, the uncertainty in the temperature is due almost entirely to the experimental uncertainty in determining $E_{\mathrm{kin,t}}$.
We note that the temperatures determined here typically lie above the estimates from the Gaussian fits of Ref.~\cite{Amerongen} ---  this illustrates  the improved  sensitivity of YY  thermometry
 and SPGPE theory.

The utility of YY thermometry  is that it relies on a single measurement---the kinetic energy per particle---which can be readily obtained via focusing and straightforward density imaging. This provides a simpler alternative to thermometry based on the measurement of density fluctuations of the gas~\cite{ARM10b}. It does not require any prior theoretical knowledge of the full momentum distribution, which is a challenging task in the strongly-correlated regime.

Finally, the results we present in this paper provide further quantitative validation of the
SPGPE in the regime of current experiments with 1D quasi-condensates (see also \cite{Cockburn2011b}).
As the SPGPE approach is suited to both equilibrium and dynamical simulations, 
this opens up an exciting avenue for exploring non-equilibrium phenomena in this system (e.g. quenches \cite{Weiler}), which cannot be explored using the YY solutions or equilibrium  quantum
Monte Carlo techniques.

In summary, we have studied the momentum properties of a finite-temperature 1D Bose gas, and given exact results for the rms width of the momentum distribution.  We have outlined a procedure from which this simple quantity can be used for sensitive kinetic-energy thermometry of a quasi-1D harmonically trapped Bose gas using the YY formalism and the LDA.  This method is applicable to all temperatures and repulsive interaction strengths.
We have also performed a full characterization of the momentum distribution in the weakly interacting regime, and made a quantitative comparison with experimental data.   Given the importance of momentum-space analysis in cold-atom research to date, our results provide a more complete picture of the homogeneous and trapped 1D Bose gas, and open up new prospects for measuring and probing these systems in the  strongly-correlated regime.

\begin{acknowledgments}
MJD~and KVK~acknowledge support by the
ARC through the 
Discovery Project scheme (DP110101047, DP1094025). PBB~acknowledges support by Marsden contract 09-UOO-0924 and FRST
contract NERF-UOOX0703. 
The work of AvA and NJvD is supported by FOM and NWO. NJvD acknowledges 
D. Stamper-Kurn for stimulating discussions.
\end{acknowledgments}

\bibliography{Ekin}

\begin{thebibliography}{42}
\bibitem [{anh()}]{anharmonicity}%
  \BibitemOpen
  \href@noop {} {}\bibinfo {note} {The slight anharmonicity of the trapping
  potential away from the trap bottom is accounted for numerically as in
  Ref.~\cite{Amerongen}.}\BibitemShut {Stop}%
\bibitem [{HF()}]{HF}%
  \BibitemOpen
  \href@noop {} {}\bibinfo {note} {The two lowest $N$ data points in
  Fig.~\ref{fig:Ekin} use $\epsilon_{\rm{cut}}=150\hbar\omega_z$, and the
  remainder with $N < 17\times \!10^3$ use
  $\epsilon_{\rm{cut}}=300\hbar\omega_z$. For $N \!\gtrsim 17\times \!10^3$,
  $\mu$ is sufficiently negative that few modes are appreciably occupied and
  the coherent region vanishes. The system is then entirely described by
  Hartree-Fock theory.}\BibitemShut {Stop}%
\end{thebibliography}

 \end{document}